\documentclass[onecolumn,secnumarabic, amssymb, noshowpacs, aps, showpacs,prb]{revtex4-1}
\usepackage{graphics}
\usepackage[dvips]{graphicx}
\usepackage{dcolumn}
\usepackage{epstopdf}
\usepackage{amsmath}
\usepackage{lipsum}
\usepackage[titletoc,toc,title]{appendix}
\usepackage{latexsym}
\usepackage[mathscr]{euscript}
\usepackage{mathrsfs}
\DeclareMathAlphabet{\mathscrbf}{OMS}{mdugm}{b}{n}
\begin{document}
\title{On the transmittance of metallic superlattices in the optical regime and the true refraction angle.}
\author{Pedro Pereyra}
\address{ Ciencias B\'{a}sicas, UAM-Azcapotzalco, M\'{e}xico D.F., M\'{e}xico}
%\date{\today}
%\maketitle
\begin{abstract}

Transmission of electromagnetic fields through $(dielectric/metallic)^n$ superlattices, for frequencies below the plasma frequency $\omega_p$, is a subtle and important topic that is reviewed and further developed here. Recently, an approach for metallic superlattices based on the finite periodic systems theory was introduced \cite{Pereyra2020}. Unlike most, if not all, of the published approaches that are valid in the $n \rightarrow \infty $ limit, the finite periodic approach is valid for any natural number $n$, and allows one to determine analytical expressions for scattering amplitudes and dispersion relations. It was shown, for frequencies below $\omega_{p}$ and large metallic-layer thickness, that under the common assumption that fields inside conductors move along the so-called "true" angle that defines the orientation of the constant-phase planes, anomalous results appear with an apparent parity effect. This issue is addressed here and it is shown that those results are due to the lack of unitarity and the underlying phenomena of absorption and loss of energy. Two compatible approaches are presented here to solve the lack of unitarity and to account for the absorption phenomenon. We show that by keeping the complex angles, the principle of flux conservation is fully satisfied, above and below $\omega_p$. The results above $\omega_p$ remain the same as in Ref. [\onlinecite{Pereyra2020}]. This approach, free of assumptions, where all the information of the scattering process is preserved, gives us light to improve the formalism when the real angle assumption is made. In fact, we show that by taking into account the induced currents and the requirement of flux conservation, we end up with an improved approach, with new Fresnel and transmission coefficients, fully compatible with those of the complex-angle approach. The improved approach allows one also to evaluate the magnitude of the induced currents and the absorbed energy, as functions of the frequency and the superlattice parameters.  We determine the plasmonic resonant frequencies, and present preliminary results of the metallic superlattices response  to electromagnetic pulses and wave packets, particularly, in the optical domain. We calculate the reflection and transmission coefficients as well as the phase time $\tau(\omega)$. We show that the predicted space-time positions agree extremely well with the actual positions of the wave-packet centroids.

\end{abstract}
%\pacs{71.70.Gm, 72.25.Hg, 72.25.Mk, 85.75.-d}
\maketitle

\section{Introduction}
The interest in the response of metallic structures to electromagnetic fields (EMFs), has grown as the possibilities of application of their properties increase. The research activity evolved along different trails, determined, by the dimension, shape, size and order of the metallic structure. The scattering of light by small metallic particles, or cylindrical and rectangular rods, require different approaches than the scattering by layered metallic structures. The geometrical differences of the scatterer systems imply, naturally, the use of different mathematical tools; For example, scattering matrices $S$, in lower dimensional cases, and the transfer matrices $M$, for layered systems.

The ancient and primordial questions on the colors of the rainbow and why the sky is blue  challenged the minds of all ages; Natural philosophers dating back to the Greeks (like Aristotle and Ptolemy), Arab physicists (like Al Kindī and Ibn Al-Haytham), and recently mighty minds like Da Vinci and Newton, explained these enigmatic phenomena, one way or another, in terms of the interaction of light and tiny particles in the atmosphere.\cite{Aristotle,Ptolemy,AlKindi,AlHazen,DaVinci,Newton} The development of electromagnetic theory led, as early as the 19th century, to the formulation of strikingly elegant, and rigorous theories on the scattering of light by small particles. In this endeavor, the seminal and influential articles by Lorenz, Rayleigh, Mie, and Debye stand out.\cite{Lorenz1880,Rayleigh,Maxwell-Garnett,Mie,Debye} The interest in the scattering of light by small particles grew rapidly after the World War II, when applied science and engineering began to produce small particles with various shapes for different purposes, usually through chemical methods. To characterize and to understand the optical properties of these systems, with more realistic shaped particles, extensive numerical methods were applied, and the Lorenz-Mie theory, for ideal spherical particles, became not only an insightful reference, but also a starting first-order approximation, in rather involved calculations; Most of the theoretical descriptions are based on numerical calculations; Countless articles have been published dealing with non-spherical particles.\cite{vanHulst1957,Kerker,vanHulst1980,Bohren,special,Haes,Kelly,Kempa,Haragushi,
Kjeldsen,Mishchenko,Gregorchuk,Ershov,Trautmann,
Zhang,Trendafilov,KJHuang} As in other fields of science, the experimental and applied research on plasmonic phenomena in metallic structures is ahead of, and move faster than, the theoretical understanding and accurate calculations, particularly, when systems contain many non-spherical particles, multiple scattering processes, shape and size dispersion and, perhaps, also the presence of random variables.

In the last thirty years, as the ability to produce low-dimensional structures grew, interest in periodic arrangements of spherical particles, cylindrical and rectangular rods, and even layered metal structures, led to the profuse field of photonic crystals. Not only does periodicity entail the possibility for simpler systems to analytically solve the light scattering problem for systems with a large number of scatterers, but introduces one of the most important known properties of periodic quantum systems, the phase coherence that is behind the band and gap structures. An important amount of properties and physics of photonic crystals, containing metallic inclusions with spherical and cylindrical symmetries,\cite{Pendry1994,BottenI2000,BottenII2000,Jacak,Bordo,Raza,Mayergoyz,Davis} have been reasonably explained, although accurate calculations are difficult to perform because of the complexity of the actual systems and of their intricate response to electromagnetic fields. Nevertheless, more accurate and appropriate theoretical descriptions are possible for layered metallic structures, as was shown in Ref. [\onlinecite{Pereyra2020}] and is further developed here.

An important class of systems with properties similar to those of the widely studied systems in the photonic crystals field, but more feasible to produce, are the flat layered $dielectric/metal$ structures, in particular periodic arrays $(dielectric/metal)^n$ where  $n$ is finite and the layers thicknesses are chosen at will.  In an attempt to study the physical properties of these structures, here called metallic superlattices, many theoretical works were published, practically all of them  assuming infinite or semi-infinite superlattices.\cite{Camley1984,Vigneron1985,Xue1985,Wallis1987,Wendler1987,Mochan1988,
Trutschel1989,Sheng1992,
Nazarov1994,Pendry1994,Quinn1995,BottenI2000,BottenII2000,Lyndin,Bria2004,Inan1999} In Ref. [\onlinecite{Pereyra2020}], a comprehensive theory was published where the finiteness is rigorously respected, and the well-known properties of electromagnetic fields inside conducting layers are taken into account.\cite{Stratton1941} Among those properties, the fact that the  wave vector ${\bf k}_c$ and the refraction angle $\theta_c$ are complex quantities, is particularly significant; the  constant-amplitude ($pz=const$) and constant-phase ($k_{cx}x+qz=const$) planes are distinct, and, as a consequence the assumption that the EMFs propagate along the so-called real or ``true" angle $\psi=\tan^{-1}(-k_{cx}/q)$ is made,  with $q$ and $p$ wave numbers as defined in equation (9). In the theoretical approach of Ref. [\onlinecite{Pereyra2020}], based on the theory of finite periodic systems, it is assumed that the electromagnetic fields, inside metals, move along the true angle $\psi$, and the transmittance as well as the plasmons' resonant frequencies are determined for almost any set of superlattice parameters, any number of unit cells, and for frequencies  above and below the plasma frequency $\omega_{p}$.\cite{PlasmaFreq} But, anomalous results were noticed for frequencies below $\omega_{p}$, and further research was offered. In this paper, we exhibit and solve the problem, and show that it is, essentially, related to the loss of flux due to neglected currents at the surfaces of the metallic layers.

When the transfer matrix that propagates EMFs across a metallic layer, of thickness $d_c$, factorizes into an attenuation factor $e^{-pd_c}$ and a matrix that accounts for the gained phases $\pm iq d_c$, the matrix becomes subunitary and the flux conservation principle is broken. The factorization, compelled by the requirement of finiteness, needs to include the surface current in order to restore the unimodularity of the transfer matrix. In this paper we face this problem and present two fully compatible approaches to deal with the transmission of EMFs through metallic layers, valid also in the optical regime. In the first approach, we deal with the complex angle and complex wave vector  assuming that the phases gained in the metallic layers are given by $\pm i{\bold k}_c\cdot {\bold d}_c$, independent of whether ${\bold k}_c$ is real or complex. In this approach, the unimodular nature of the transfer matrices, thus the principle of flux conservation, is rigorously preserved. In the second approach, we consider again the real-angle approach but include now the induced currents at the metallic layers and impose a flux conservation requirement. This allows us to determine the magnitude of the induced currents and to define an absorption factor $a$. The calculation of the transmittance shows that the predictions of both approaches agree completely. An advantage of the second approach is that it allows to obtain an insight on the induced currents and the absorption factor.

In Section 2, we review the boundary conditions and show the origin of the lack of unitarity in the transfer matrices. We show that, for frequencies below the plasma frequency $\omega_{p}$, the assumption of EMFs moving along the true angle and neglecting, at the same time, the induced currents (in the metallic layers), lead to a lack of unitarity. In section 3 we present the complex angle approach (CAA) and show that, applying the TFPS without being disturbed by the complex nature of the refracted angle,  i. e. without any reference to the constant phase and constant plane projections, we obtain an approach that works well below and above $\omega_{p}$. In section 4 we turn into the true angle approach but now taking into account in the boundary conditions the currents induced by the electric polarization of the right and left moving fields and, on top of this,  we impose the flux conservation requirement in a way to include the absorbed energy. This approach led us not only to determine the magnitude and phase of the effective induced currents, but also to present a unified and more insightful approach that works equally well for frequencies below and above the plasma frequency $\omega_{p}$. The explicit transmittance calculations show that, in the low frequencies ($\omega < \omega_{p}$) domain, this approach agrees with the complex angle approach predictions. In the  high frequencies ($\omega > \omega_{p}$) domain, both approaches coincide with the predictions of the approach in Ref. [\onlinecite{Pereyra2020}].

We will see that in the domain of frequencies $\omega < \omega_{p}$, where the reflection is large, a photonic band structure, with almost complete transmission, emerges, when the dielectric width $d_a$ increases. We will see also that the low frequency resonances describe highly localized polarons with large mean-life time. As in Ref. [\onlinecite{Pereyra2020}], qualitative and quantitative differences are observed in the transmittance features above and below $\omega_{p}$, with strong dependence on the incidence angle and the superlattice parameters. We will see that the resonant dispersion relation, derived in the TFPS, predicts the band widths and the frequencies of the plasmonic resonances.

In this paper we present also results of reflected, transmitted and tunneling times of  Gaussian electromagnetic wave packets by metallic superlattices (MSLs). A detailed analysis of the space-time evolution of Gaussian wave packets will be publish elsewhere.

\section{Unitarity deficit in the constant-phase direction}

\begin{figure}
\begin{center}
\includegraphics [width=10.6cm]{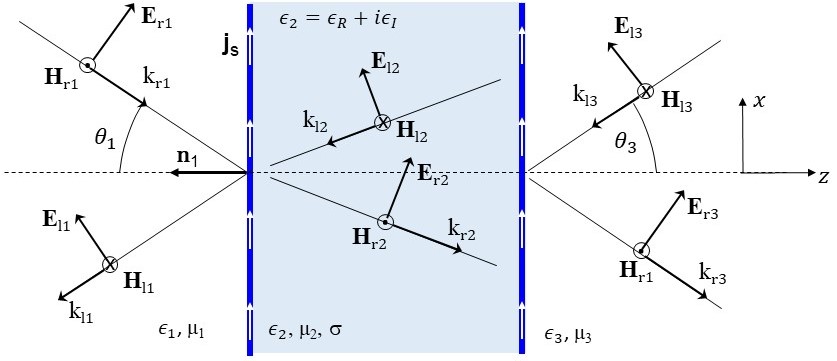}
\caption{The right and left moving electromagnetic waves in the superlattice layers.}\label{f1}
\end{center}
\end{figure}

In this section we will show that the assumption of EMFs moving along the true angle $\psi$ and the requirement of finite EMFs, lead to break the principle of flux conservation, for frequencies below the plasma frequency $\omega_{p}$. To make clear this effect, let us assume an electromagnetic field with, say, parallel polarization and incidence angle $\theta_j$, moving across a superlattice $(D_1/M_2/D_3)^n$, where $n$ is the number of unit cells. The unit cell $D_1/M_2/D_3$ comprises two dielectric ($D_j$) layers characterized by electric permittivities $\epsilon_j$ and magnetic permeabilities $\mu_j$, and a metal ($M_2$) layer, with dielectric function written as\cite{Yang}
\begin{equation}
\epsilon_2(\omega)=\epsilon_{\infty}-\frac{E_p^2}{\hbar^2\omega^2}+i\frac{\sigma}{\omega},
\end{equation}
For specific calculations we will consider silver parameters: $\epsilon_{\infty}=5.7$ and $E_p= 9eV$, with plasma frequency\cite{PlasmaFreq,Pendry2000} $\omega_p=E_p/(\hbar\sqrt{\epsilon_{\infty}})\simeq 5.729\times 10^{15}s^{-1}$. We will assume also that $\mu_1=\mu_3=\mu_2=1$. Regardless of whether the electric parameters are real or complex, the electric and magnetic fields in layer $j$, see figure \ref{f1}, can be written as
\begin{eqnarray}
  {\bold E}_j(r,t) &=&  {\bold E}_{rj}+{\bold E}_{lj}= \mathcal{E}_{rj} {\bold n}_{rj} e^{i ({\bold k}_{rj}\cdot {\bold r}-\omega t)}+ \mathcal{E}_{lj}{\bold n}_{lj} e^{-i ({\bold k}_{lj}\cdot {\bold r}+\omega t)}\hspace{0.3in} {\rm for}\hspace{0.3in} j=1,2,3 \cr
   {\bold H}_j(r,t) &=& {\bold H}_{rj}+{\bold H}_{lj}= \frac{{\bold k}_{rj}\times {\bold n}_{rj}}{\omega \mu_j} \mathcal{E}_{rj}  e^{i ({\bold k}_{rj}\cdot {\bold r}-\omega t)}+ \frac{{\bold k}_{lj}\times {\bold n}_{lj}}{\omega \mu_j}  \mathcal{E}_{lj}{\bold n}_le^{-i ({\bold k}_{lj}\cdot {\bold r}+\omega t)}
\end{eqnarray}
where  ${\bold n}_{rj}$ and ${\bold n}_{lj}$ are the polarization vectors
\begin{equation}\label{PolarizVectors}
  {\bold n}_{rj}=(\cos \theta_j,0,\sin \theta_j)\hspace{0.3in}{\rm and}\hspace{0.3in}{\bold n}_{lj}=(\cos \theta_j,0,-\sin \theta_j),
\end{equation}
of the right and left moving fields, and the corresponding wave vectors are
\begin{equation}\label{wavevectors}
{\bold k}_{rj}=k_{j} {\bold u}_{rj}=\omega \sqrt{\mu_j\epsilon_j} (-\sin \theta_j,0, \cos \theta_j)\hspace{0.3in}{\rm and}\hspace{0.3in}{\bold k}_{lj}=k_{j} {\bold u}_{lj}=\omega \sqrt{\mu_j\epsilon_j} (-\sin \theta_j,0,- \cos \theta_j).
\end{equation}
For a simplified notation, we define the field phases
\begin{eqnarray}
\phi_{rj}&=&k_{rjx}x+k_{rjz}z-\omega t=-k_j x \sin \theta_j+k_j z \cos \theta_j -\omega t \cr \phi_{lj}&=&k_{ljx}x-k_{ljz}z-\omega t=-k_j x \sin \theta_j - k_j z \cos \theta_j -\omega t.
\end{eqnarray}
Thus, the field components in each layer $j$-th of the superlattice are
\begin{eqnarray}
  E_{jx} &=& (\mathcal{E}_{rj}e^{i \phi_{rj}}+\mathcal{E}_{lj}e^{i \phi_{lj}})\cos \theta_j \crcr
   E_{jz} &=& (\mathcal{E}_{rj}e^{i \phi_{rj}}-\mathcal{E}_{lj}e^{i \phi_{lj}})\sin \theta_j\crcr
   H_{jy} &=& \frac{k_j}{\omega \mu_j} (\mathcal{E}_{rj}e^{i \phi_{rj}}-\mathcal{E}_{lj}e^{i \phi_{lj}}).
\end{eqnarray}
\begin{figure}
\begin{center}
\includegraphics [width=10.6cm]{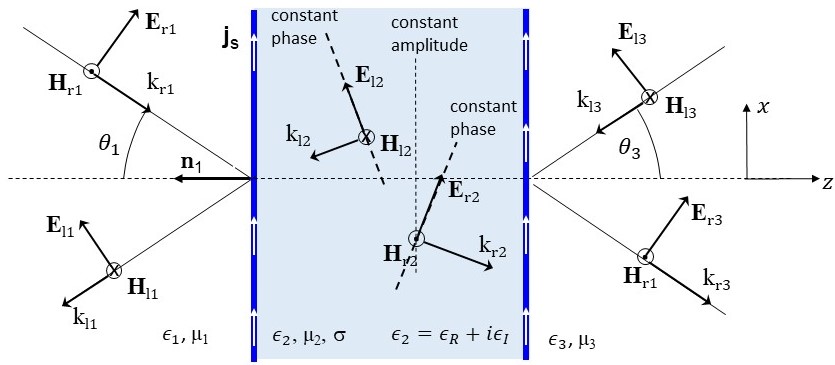}
\caption{Constant-phase and constant-amplitude planes for the electromagnetic fields propagating in the conductor media.}\label{f2}
\end{center}
\end{figure}
Applying the boundary condition for the electric fields at the interface $1|2$, we obtain the  the well-known Snell's law
\begin{equation}\label{Snell}
  k_2\sin \theta_2=k_1 \sin \theta_1,
\end{equation}
with $k_2$ and $\sin \theta_2$ complex. Because of the complex nature of these quantities, it has been  common to represent the electromagnetic field phases in the conductor as\cite{Stratton1941}
\begin{eqnarray}
\phi_{r2}&=&(k_{2R}+i k_{2I})(- x \sin \theta_2+ z \cos \theta_2) -\omega t= a_2 x +q z +i( b_2 x +p z) -\omega t \cr \phi_{l2}&=&(k_{2R}+i k_{2I})(- x \sin \theta_2- z \cos \theta_2) -\omega t=a_2 x - q z +i( b_2 x - p z) -\omega t.
\end{eqnarray}
Here $k_{2x}=a_2+i b_2=-k_1 \sin \theta_1$, and $k_{2z}=q+ip$, with
\begin{eqnarray}
q=\rho \left(k_{2R}\cos \frac{\chi}{2}-k_{2I}\sin\frac{\chi}{2} \right) \hspace{0.3in}{\rm and}\hspace{0.3in}
p=\rho \left(k_{2I}\cos \frac{\chi}{2}+k_{2R}\sin\frac{\chi}{2}\right),
\end{eqnarray}
and
\begin{eqnarray}
\rho=\left(1+\frac{2 k_1^2(k^2_{2I}-k^2_{2R})\sin^2\theta_1+k^4_1 \sin^4 \theta_1}{k_2^2}\right)
\hspace{0.3in}{\rm and}\hspace{0.3in}
\chi=\tan^{-1}\left( \frac{2 k_1^2 k_{2I} k_{2R} \sin^2 \theta_1}{k_2^2+2 k_1^2(k^2_{2I}-k^2_{2R})\sin^2 \theta _1} \right).
\end{eqnarray}
To recognize that, in the propagation of the electromagnetic waves inside conductors, one can distinguish directions of constant-phase and directions of constant-amplitude, as sketched in figure \ref{f2}.  The constant-phase and constant-amplitude planes, defined by
\begin{equation}\label{ConstantPhase}
 a_2 x \pm q z= constant \hspace{0.3in}{\rm and}\hspace{0.3in}  b_2 x \pm p z = constant,
\end{equation}
respectively, propagate along the normals making angles $\psi$ and $\psi'$ with the $z$-axis, defined by
\begin{equation}
\tan \psi=-\frac{a_2}{q} \hspace{0.3in}{\rm and}\hspace{0.3in} \tan \psi'=-\frac{b_2}{p},
\end{equation}
respectively. An important and well-known consequence of the complex phases is the attenuation of the electromagnetic fields, with the aftermath of loss of energy through the longitudinal and transverse currents induced by the electric fields. The rather common assumption made in Ref. [\onlinecite{Pereyra2020}] that the electromagnetic fields move along the direction determined by the true or real angle $\psi$,\cite{Inan1999} neglecting the  longitudinal and transverse currents, supposedly concentrated only at the surfaces, led one to write the transfer matrix that connects field vectors at the left and right of the conductor as
\begin{equation}\label{UnitCellTrueAngleTM}
M'_c=\frac{1}{2 \kappa\mu_1 \cos\psi+2\xi}\left( \begin{array}{cc} \alpha_l & \beta_l \cr \beta_l^* & \alpha_l^* \end{array}\right)\left( \begin{array}{cc} e^{i( q+i p) d_c} & 0 \cr 0 & e^{-i( q -i p)d_c} \end{array}\right)\frac{1}{2 k_1\mu_2 \cos\theta_1}\left( \begin{array}{cc} \alpha_l^* & -\beta_l \cr -\beta_l^* & \alpha_l \end{array}\right),
\end{equation}
$p$ and $q$ are as defined before, $\kappa=\bigl(q^2+k_1^2\sin^2\theta_i \bigr)^{1/2}$, $\xi=k_1\mu_1\sec\theta_1 \tan\psi$ and
\begin{eqnarray}
 \alpha_l&=&  k_1\mu_2 \sec\theta_1 +\kappa \mu_1 \cos\psi+ \xi+i\, p\mu_1 \cr
 \beta_l &=& k_1\mu_2 \sec\theta_1-\kappa \mu_1 \cos\psi-\xi+i\, p\mu_1.
\end{eqnarray}
Note that, to keep the physical quantities finite, one of the important matrices in (\ref{UnitCellTrueAngleTM}), the transfer matrix that connects the fields at the beginning and the end of the metallic layer, was written as
\begin{equation}\label{finitephase}
  M_m=\left( \begin{array}{cc}
               e^{i(q+ip)d_c} & 0 \\
               0 & e^{-i(q-ip)d_c}
             \end{array} \right),
\end{equation}
with a phase $-i(q-ip)d_c$ instead of $-i(q+ip)d_c$ for the left moving EMF.
\begin{figure}
\begin{center}
\includegraphics [width=15.4cm]{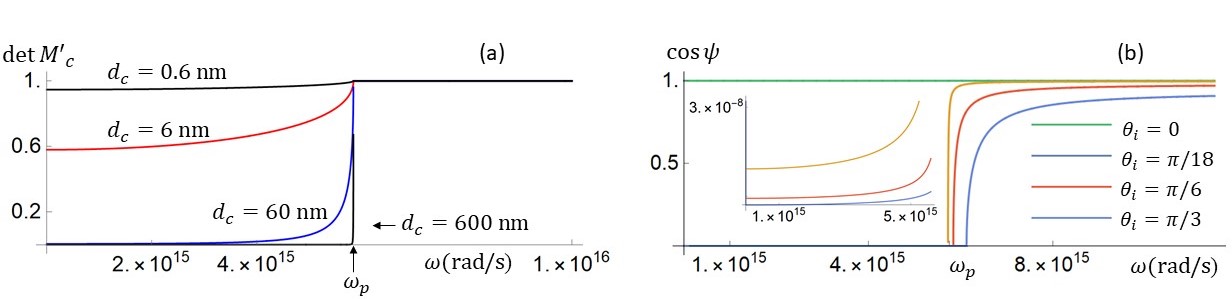}
\caption{Determinant of the transfer matrix $M'_c$ in (\ref{UnitCellTrueAngleTM}), and the real angle as functions of frequency, conductor-layer width and incidence angle. (a) The determinant of the unit-cell transfer matrix, $\det M'_c$, decreases as the conductor layer width $d_c$ increases. (b) The cosine of the real angle $\psi$. We show here that even though that for $\omega>\omega_p$ it is close to 1, as is well known, for $\omega<\omega_p$ is close to zero, except for $\theta_i=0$.  }\label{f3}
\end{center}
\end{figure}
A simple way to have a measure of the lack of flux is the unimodularity of the transfer matrix. In Figure \ref{f3} (a), we plot the determinant of the unit-cell transfer matrix $M'$, shown in Eq. (\ref{UnitCellTrueAngleTM}) and used in Ref. [\onlinecite{Pereyra2020}], as a function of frequency $\omega$, and for different values of the conducting layer width $d_c$. It is clear in this figure that, for frequencies below $\omega_p$,  the determinant of $M'$ is less than 1 and decreases as the layer width $d_c$ increases. Another quantity whose behavior needs to be clarified is the real or true angle $\psi$ that defines the direction of constant-phase planes. It is generally assumed that its values, measured from the normal to the interface, are small. That is true for frequencies above $\omega_p$. But, that is not true for $\omega<\omega_p$, as shown in Figure \ref{f3} (b). Except for $\theta_i=0$, where $\tan \psi$ is rigorously zero, $\psi$ is close to $\pi/2$ for $\theta_i\neq 0$ and $\omega <\omega_p$, as shown in the inset of the figure. This means that when the electric field has a large component parallel to the interface, the planes of constant-phase tend to propagate parallel to the interface, for small frequencies and finite conductivity.

Our purpose in the next sections is to solve the lack of unitarity in the optical regime. We will show that both the finiteness and the lack of unitarity can be overcome. We will present  two approaches, with compatible results. We shall first consider a complex-angle approach, and  in section \ref{sec4}, we will present an improved real-angle approach, where, besides the true angle assumption, we will take into account, explicitly, the induced currents. In the complex-angle approach, we keep the complex angle and the complex wave number at the conducting layers. In this case, the unit-cell transfer matrix is automatically unimodular. In the improved real-angle approach, the induced currents account for the absorption factor and restore the flux-conservation principle.

\section{Complex-angle approach}

Even though the recognition of constant-phase and constant-amplitude planes for EMFs inside conductors helps one to understand important properties of these fields in this kind of media, there is no need, in principle, in  the transfer matrix approach to assume that the EMWs move along the real or true angle. In fact, we can equally well work with the complex angle $\theta_2$, be compatible with the principle of flux conservation, and we can determine the transmission and reflection coefficients of metallic superlattices for frequencies below and above the plasma frequency $\omega_p$. In this section, we will present this complex-angle approach, and obtain some results.

Te boundary conditions at the interface $1|2$, at $z=z_l$, are expressed by the following relation
\begin{equation}\label{TransicMatrix}
\left(\begin{array}{c}
  E_{r2x}\cr
  E_{l2x} \end{array}\right)_{z_l^+}=\left(\begin{array}{c}
  \mathcal{E}_{r2}e^{i\phi _{r2}}\cos \theta_2\cr
   \mathcal{E}_{l2}e^{i\phi _{l2}}\cos \theta_2 \end{array}\right)_{z_l^+}=\frac{1}{2} \left(\begin{array}{cc}
    1+B_l & 1-B_l \cr
     1-B_l & 1+B_l
        \end{array} \right) \left(\begin{array}{c}
  \mathcal{E}_{r1}e^{i\phi _{r1}}\cos \theta_1\cr
   \mathcal{E}_{l1}e^{i\phi _{r1}}\cos \theta_1\end{array}\right)_{z_l^-}=M_l(z_l^+,z_l^-)\left(\begin{array}{c}
  E_{r1x}\cr
  E_{l1x} \end{array}\right)_{z_l^-}
\end{equation}
where the subindices $z_l^-$ and $z_l^+$ mean evaluation at $z_l-\epsilon$ and $z_l+\epsilon$, in the $\lim \epsilon\rightarrow 0$. The parameters $B_l$ and $cos \theta_2$ are
\begin{equation}
B_l=\frac{k_1\mu_2  \cos \theta_2}{ k_2\mu_1\cos \theta_1}\hspace{0.3in}{\rm and}\hspace{0.3in}\cos \theta_2=\rho e^{i\, \chi/2}
\end{equation}
Similarly, the boundary conditions at the interface $2|3$, where $z=z_r$, lead to
\begin{equation}\label{TransicMatrix}
\left(\begin{array}{c}
  E_{r3x}\cr
   E_{l3x}\end{array}\right)_{z_r^+}=\frac{1}{2} \left(\begin{array}{cc}
    1+B_r & 1-B_r \cr
     1-B_r & 1+B_r
        \end{array} \right) \left(\begin{array}{c}
  E_{r2x}\cr
   E_{l2x}\end{array}\right)_{z_r^-}=M_r(z_r^+,z_r^-)\left(\begin{array}{c}
  E_{r2x}\cr
   E_{l2x}\end{array}\right)_{z_r^-}.
\end{equation}
Here
\begin{equation}
B_r=\frac{k_2\mu_1  \cos \theta_1}{ k_1\mu_2\cos \theta_2}.
\end{equation}
The matrices $M_l$ and $M_r$ are formally similar to the well-known matrices of Fresnel amplitudes. In Ref. [\onlinecite{Pereyra2020}], we had $\psi$ instead of $\theta_2$. In the following, we will omit the subindices $z_x^{\pm}$ for the matrices. The relation between the electromagnetic fields at the left and right hand side of the conducting layer is
\begin{equation}\label{TransicMatrix}
\left(\begin{array}{c}
  E_{r3x}\cr
   E_{l3x}\end{array}\right)=\frac{1}{2} \left(\begin{array}{cc}
    1+B_r & 1-B_r \cr
     1-B_r & 1+B_r
        \end{array} \right)\left(\begin{array}{cc}
    e^{i\,\varphi_c} & 0 \cr
    0 &  e^{i\,\varphi_c}
        \end{array} \right)\frac{1}{2}\left(\begin{array}{cc}
    1+B_l & 1-B_l \cr
     1-B_l & 1+B_l
        \end{array} \right) \left(\begin{array}{c}
  E_{r1x}\cr
   E_{l1x}\end{array}\right)=M_c\left(\begin{array}{c}
  E_{r1x}\cr
   E_{l1x}\end{array}\right),
\end{equation}
where $\varphi_c=k_2 d_c \cos \theta_2$, when the conducting layer width is $d_c$. Multiplying the matrices we obtain the transfer matrix for a conductor layer
\begin{equation}
M_c=\left(\begin{array}{cc} \alpha_c & \beta_c \\
                                        \gamma_c & \delta_c
                                      \end{array}  \right),
\end{equation}
with
\begin{eqnarray}
 \alpha_c &=& \cos \varphi_c +i\displaystyle{\frac{k_1^2\cos^2 \theta_2+k_2^2\cos^2 \theta_1}{2 k_1 k_2 \cos \theta_1 \cos \theta_2}}\sin \varphi_c  \hspace{0.6in}\beta_c= i\displaystyle{\frac{k_2^2\cos^2 \theta_1-k_1^2\cos^2 \theta_2}{2 k_1 k_2 \cos \theta_1 \cos \theta_2}}\sin \varphi_c \cr \vspace{0.2in}& & \cr
  \delta_c &=& \cos \varphi_c -i\displaystyle{\frac{k_1^2\cos^2 \theta_2+k_2^2\cos^2 \theta_1}{2 k_1 k_2 \cos \theta_1 \cos \theta_2}}\sin \varphi_c  \hspace{0.6in}\gamma_c= -i\displaystyle{\frac{k_2^2\cos^2 \theta_1-k_1^2\cos^2 \theta_2}{2 k_1 k_2 \cos \theta_1 \cos \theta_2}}\sin \varphi_c.
\end{eqnarray}
Therefore, the transfer matrix of the unit cell $D_1/M_2/D_3$, whose layer widths are $d_a/2$, $d_c$ and $d_a//2$, is
\begin{eqnarray}
M=\left(\begin{array}{cc}
          e^{i\,\theta_a/2}& 0 \\
          0 & e^{-i\,\theta_a/2}
        \end{array}\right) \left(\begin{array}{cc}  \alpha_c & \beta_c \\
                                        \gamma_c & \delta_c
                                      \end{array}  \right)\left(\begin{array}{cc}
          e^{i\,\theta_a/2}& 0 \\
          0 & e^{-i\,\theta_a/2}
        \end{array}\right)=\left(\begin{array}{cc} \alpha & \beta\\
                                        \beta^* & \alpha^*
                                      \end{array}  \right)
\end{eqnarray}
Here
\begin{equation}
\theta_a=d_ak_1 \cos \theta_1.
\end{equation}
It is worth noticing that in this representation, both flux and  time reversal invariance  are preserved. This is so because keeping the complex angle, all the components of the electromagnetic field, i.e. the transmitted, reflected and absorbed components, are fully taken into account. As will be seen below, the transmission coefficients in the high frequency domain, for  $\omega > \omega_p$ are exactly the same as in the real-angle approach. However, in the optical regime, where $\omega < \omega_p$, some properties remain, but others like the apparent parity effect and transparency, disappear.

\begin{figure}
\begin{center}
\includegraphics [width=15.6cm]{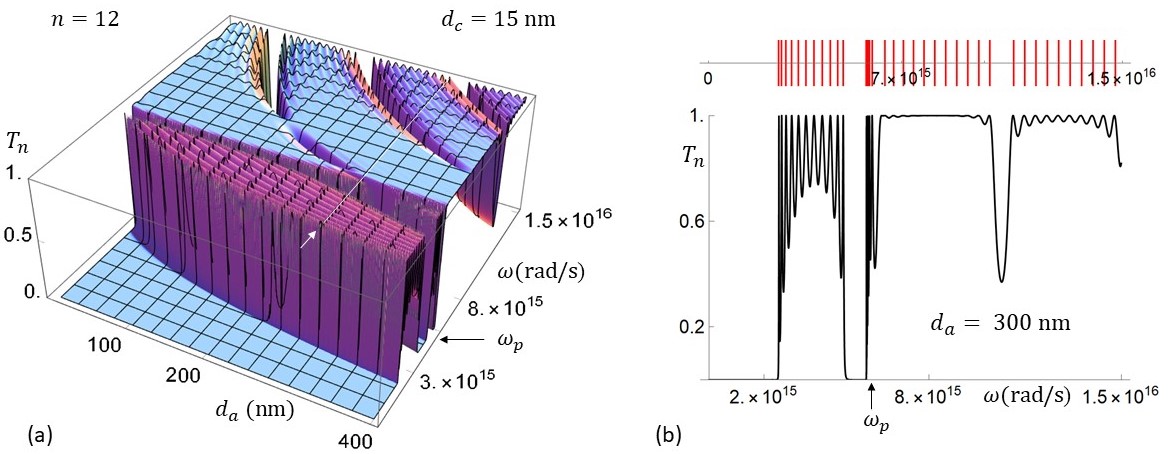}
\caption{Transmission coefficient as a function of $\omega$ and $d_a$, for a metallic superlattice with $n=12$, $d_c=15$nm and $\theta_i=\pi/3$. In (a) the bands move to lower frequencies as $d_a$ grows. In (b) we see more clearly the resonant transmission for a fix value of the conductor layer width $d_c$. In the upper part of (b), bands and resonances predicted by the resonant dispersion relation (\ref{RDR}).}\label{f4}
\end{center}
\end{figure}
As was pointed out in Ref. [\onlinecite{Pereyra2020}], to determine the scattering properties of the electromagnetic fields through metallic superlattices, using the theory of finite periodic systems\cite{Pereyra1998,Pereyra2002,Pereyra2005,Simanjuntak2007,Pereyra2008,Pereyra2012} outlined there, it is essential to know the unit-cell transfer matrix, and applying the general formulas of the TFPS one can, straightforwardly, obtain the transmittance and reflectance through metallic superlattices, as well as to determine the resonant band structure of the surface plasmon polaritons. The transmission and reflection coefficients of a superlattice with $n$ unit cells, are obtained from
\begin{equation}\label{TransmTn}
  T_n=\frac{1}{|\alpha_n|^2}  \hspace{0.3in}{\rm and}\hspace{0.3in}   R_n=\frac{|\beta_n|^2}{|\alpha_n|^2},
\end{equation}
where
\begin{equation}\label{alfan}
  \alpha_n=U_n-\alpha^* U_{n-1}  \hspace{0.3in}{\rm and}\hspace{0.3in} \beta_n = \beta^{-1}U_{n-1},
\end{equation}
and $U_n$ the Chebyshev polynomial of the second kind and order $n$, evaluated at the real part of $\alpha$. The resonant frequencies and band widths are determined by the resonant dispersion relation\cite{Pereyra2002,Pereyra2005,Pereyra2017,Pereyra2020}
\begin{equation}
\cos \frac{\nu+(\mu-1)n}{n}\pi=(\alpha_R)_{\mu,\nu}\hspace{0.3in} {\rm with} \hspace{0.3in}\mu=1,2,3,...  \;\; \nu=1,2,...,n-1.
\end{equation}
$\mu$ and $\nu$ are the quantum numbers of the resonant frequencies $\omega_{\mu,\nu}$, of the $\nu$-th resonance of the band $\mu$. Generally $\mu=$1, 2, 3, ... and $\nu=$1, 2, ..., $n$-1.

In terms of the physical quantities defined in this approach, the resonant dispersion relation is

\begin{eqnarray}\label{RDR}
\cos \frac{\nu \!+\!(\mu\!-\!1)n}{n}\pi\!=Re\left[(\cos 2\theta_a+i \sin 2\theta_a)\left(\cos \varphi_c+i\frac{k_1^2 \cos^2\theta_2+k_2^2 \cos^2\theta_1}{2k_1k_2 \cos \theta_1 \cos \theta_2}\sin \varphi_c\right)\right]_{\omega_{\mu,\nu}}
\end{eqnarray}

\begin{figure}
\begin{center}
\includegraphics [width=15.6cm]{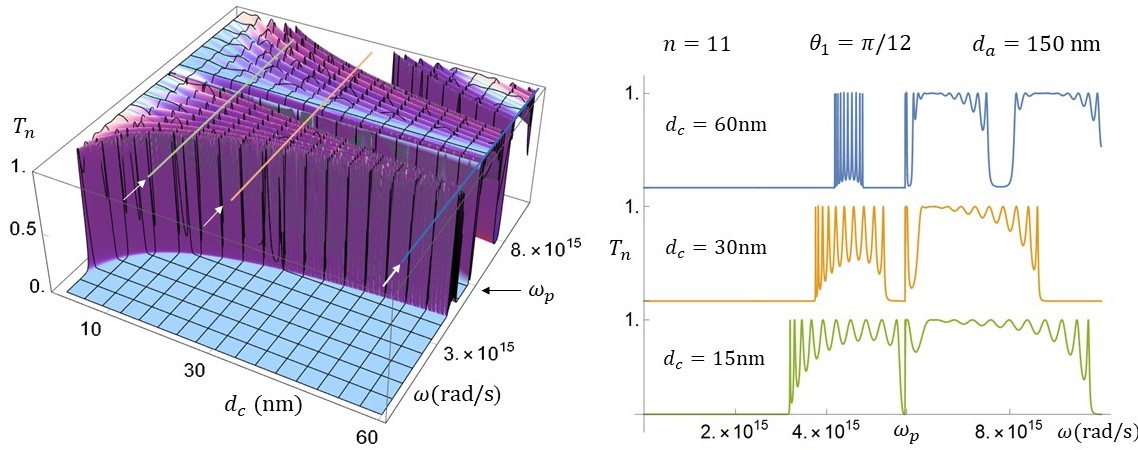}
\caption{Transmission coefficient as a function of $\omega$ and $d_c$, for a metallic superlattice with $n=11$, $d_a=150$nm and $\theta_i=\pi/12$.  In (a), the band width diminishes as the conductor layer width $d_c$ increases. This effect is shown in (b) for three values of $d_c$, indicated with white arrow in (a).}\label{f5}
\end{center}
\end{figure}

In Figures \ref{f4} and \ref{f5} we show the trends of the transmission coefficient as function of the frequency and of the layer widths. The frequencies vary from 0 to $1.5\times10^{16}$Hz, i.e. frequencies below and above $\omega_p$. In Figure \ref{f4}, the transmission is plotted as a function of the dielectric-layer width $d_a$, while in Figure \ref{f5} as a function of the conducting layer width $d_c$. For these examples and the others in this report, we consider air in the dielectric layers and silver in the metallic ones.

For frequencies above $\omega_p$, which for silver is of the order of $5.72\times10^{15}$Hz, the transmission coefficients are, in all cases, exactly the same as those in the real-angle approach of Ref. [\onlinecite{Pereyra2020}], however, for frequencies below $\omega_p$, the results are different. Below the plasma frequency, we have now narrower bands and thinner resonant states, implying larger mean-life times for the resonant states and larger tunneling times for electromagnetic waves whose frequencies are resonant. In  electronic and electromagnetic field transport, the transmission resonances result from a complex and coherent superposition of electromagnetic fields facilitated by the superlattice periodicity. These coherent superposition of fields imply the participation of collective photon-driven electron oscillations, the so-called plasmon polaritons.  The long standing resonances correspond to localized plasmon polaritons. From figures \ref{f4} and \ref{f5} it is clear  that the increasing of $d_a$ and the increasing of $d_c$ have opposite effects on the extension of the complete-reflection domain, at low frequencies. While increasing $d_a$ the reflection domain diminishes, increasing $d_c$ the complete-reflection domain grows.

In Figures \ref{f4} (a) and (b), we have the transmission coefficient as a function of $\omega$ and $d_a$, for a metallic superlattice with $n=12$, $d_c=15$nm and $\theta_i=\pi/3$. In Figure \ref{f4} (a), we see that increasing $d_a$ the bands move to lower frequencies, as in quantum systems when quantum-well widths increase, and the band widths become narrower. To visualize the resonant-band features better, we plot in Figure \ref{f4} (b) the transmission coefficient evaluated at $d_a=300$nm, indicated with the white arrow in (a). We also plot, in the upper part of this graph, the resonant levels predicted by Eq. (\ref{RDR}). These resonances, that result from complex coherent superpositions of multiple reflected and transmitted fields plus collective electron oscillations are, in generally, extended electromagnetic states, with large mean-life time, nevertheless the collective electron oscillations occur mainly at the surfaces of the metallic layers. These surface excitations correspond to the so-called localized surface plasmon polaritons.

\begin{figure}
\begin{center}
\includegraphics [width=15.6cm]{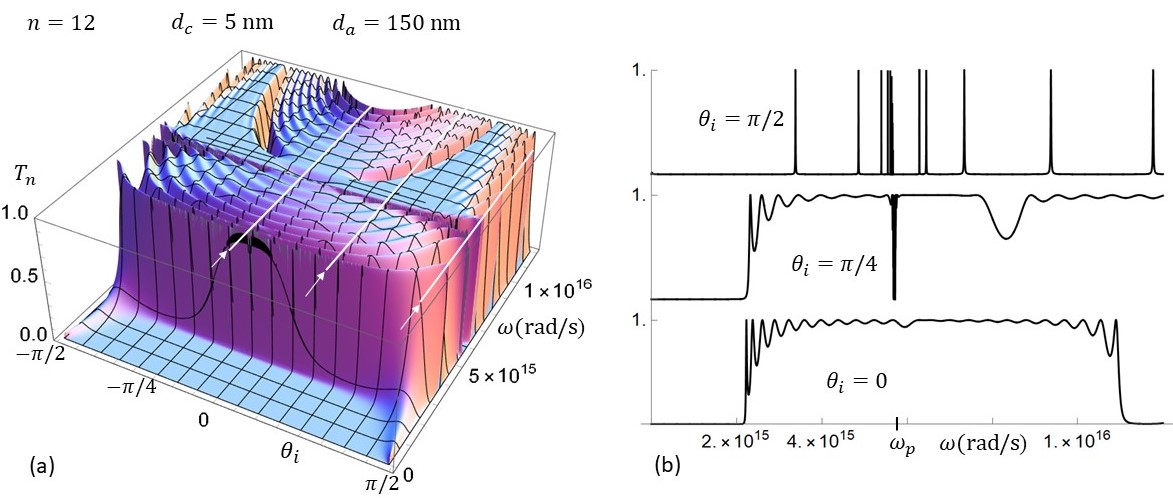}
\caption{Transmission coefficient as a function of  $\omega$ and $\theta_i$, for a metallic superlattice with $n=12$, $d_a=150$nm and $d_c=5$nm. In (b), the transmission coefficient for  $\theta_i=0$, $\pi/4$ and $pi/2$. These graphs correspond to those in (a) indicated with white arrows.}\label{f6}
\end{center}
\end{figure}
In Figures \ref{f5} (a) and (b), we have the transmission coefficient as a function of $\omega$ and $d_c$, for a metallic superlattice with $n=11$, $d_a=150$nm and $\theta_i=\pi/12$. In this example, we have also a resonant transmission band for $\omega < \omega_p$. In Figures \ref{f5} (a), we see that the band width diminishes as the conductor layer increases. To visualize the resonant behavior better, we plot in Figure \ref{f5} (b) the transmission coefficients for $d_c=15$nm, $d_c=30$nm and $d_c=60$nm. They correspond to those layers' widths indicated with white arrows in Figure \ref{f5} (a). Here also the resonant transmission bands are within the visible light domain.

The resonances positions and band widths depend not only on the layers' widths, but also depend strongly on the incidence angle. In figures \ref{f6} (a) and (b), we plot the transmission coefficient as a function of  $\omega$ and $\theta_i$, for a metallic superlattice with $n=12$, $d_a=150$nm and $d_c=5$nm. In Figure \ref{f6} (b), we see the transmission coefficient for  $\theta_i=0$, $\pi/4$ and $pi/2$, which correspond to those values of $\theta_i$ indicated with white arrows in (a). It is clear from these graphs the enormous qualitative and quantitative differences in the transmission coefficient and the resonant  features as functions of the incidence angle. As was shown in Ref. [\onlinecite{Pereyra2020}] the transmission resonances become delta type when the incidence angle is equal or even close to $\pi/2$.
\begin{figure}
\begin{center}
\includegraphics [width=15.6cm]{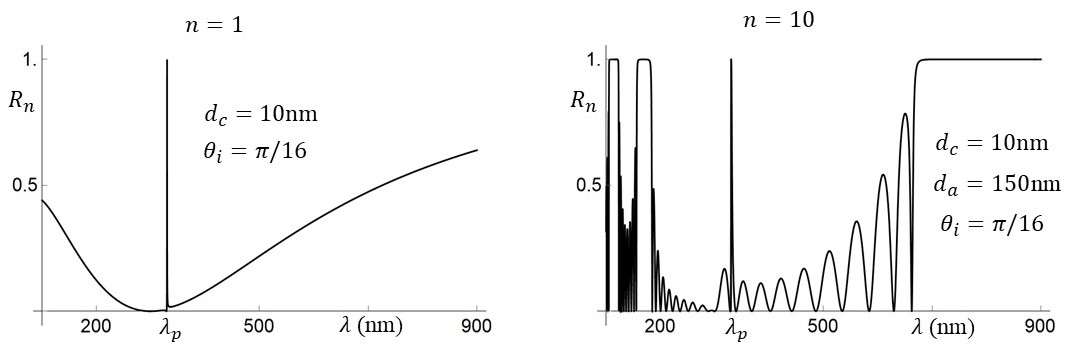}
\caption{The reflection coefficient $R_n$ as a functions of the wavelength $\lambda$. In (a) for a single layer of silver. In (b) for a metallic superlattice with $n=10$. At the plasma wavelength $\lambda_p$, we see a characteristic Fano-like resonance, due to the
interference between propagating and evanescent modes.}\label{f7}
\end{center}
\end{figure}
Since $R_n=1-T_n$, we generally omit the calculation of the reflection coefficient. Nevertheless, it may be helpful to visualize its behavior, in particular, the effect that the number of layers may have in the optical regime. In Figure \ref{f7}, we have the reflection coefficient of a single silver layer (a) and for a superlattice with $n=10$. It is clear that the superlattice not only imply resonances, but also well defined bands of complete reflection and, for some configurations, of complete transmission.

Playing with the superlattice parameters, the incidence angle and the electromagnetic field polarization, we can obtain an endless variety of optical responses for the electromagnetic fields that fall upon the surface of a metallic superlattice. In the next section we will return to the real angle approach and compare with the results of this approach.

\section{Improved real-angle approach}\label{sec4}

As explained before, the lack of flux in the real-angle approach, indicates the need to include the induced currents at the metallic layers. We will include the induced currents, assume that inside the conductor the electromagnetic waves move along the real angle (also called true angle) of the constant-phase planes, and finally impose the conservation of flux. It is well known that the inclusion of currents, induced in the metallic layers and responsible of the plasmonic resonances, is compatible with the Maxwell equation
\begin{equation}\label{MaxwellEq}
  {\bold \nabla}\times {\bold H}={\bold J}+\epsilon\frac{\partial \bold E}{\partial t}
\end{equation}
which contemplates the possibility of induced currents, the plasmon oscillations, which in perfect conductors are localized at the surface, but in lossy conductors, respond to the local field. We will assume that in each layer $j$ we have currents, induced be the right and left moving fields, whose magnitudes are proportional to the magnitude of the local electric fields
\begin{equation}\label{currents}
  J_{rj}(x,z)= s_R {E}_{rj}(x,z) \hspace{0.2in}{\rm and}\hspace{0.2in}J_{lj}(x,z)= s_L {E}_{lj}(x,z),
\end{equation}
Here the amplitudes  $s_L$ and $s_R$ are proportional to the conductivity $\sigma$ and  to an attenuation factor determined below, that accounts for the loss of energy. Taking into account the surface currents
\begin{eqnarray}\label{surfacecurrents}
J_{rxl}=s_{R}{E}_{r2}(x,z_{l})\cos\psi =s_{R}{E}_{r2x}\big|_{z_l},\hspace{0.3in}{\rm and}\hspace{0.3in} J_{lxl}=s_{L}{E}_{l2}(x,z_{l})\cos\psi =s_{L}{E}_{l2x}\big|_{z_l},
\end{eqnarray}
at $z=z_l$, and
\begin{eqnarray}
J_{rxr}=s_{R}{E}_{r2}(x,z_{r})\cos\psi =s_{R}{E}_{r2x}\big|_{z_r},\hspace{0.3in}{\rm and}\hspace{0.3in}
J_{lxr}=s_{L}{E}_{l2}(x,z_{r})\cos\psi =s_{L}{E}_{l2x}\big|_{z_r},
\end{eqnarray}
at $z=z_r$, for the boundary conditions. The transfer matrix of the conducting layer that connects the field vectors at its left and right hand sides becomes
\begin{equation}
M_{\jmath c}=\frac{1}{2 \kappa\mu_1 \cos\psi+2\xi}\left( \begin{array}{cc} \alpha_l-s_R & \beta_l-s_L \cr \beta_l^*+s_R & \alpha_l^*+s_L \end{array}\right)M_m(z_r^-,z_l^+) \frac{1}{2 k_1\mu_2 \cos\theta_1}\left( \begin{array}{cc} \alpha_l^* -s_L& -\beta_l -s_L\cr -\beta_l^*+s_R & \alpha_l+s_R \end{array}\right),
\end{equation}
with  $p=\rho (\epsilon_R \sin \gamma+\epsilon_I \cos \gamma)$, $\kappa=\bigl(q^2+k_1^2\sin^2\theta_i \bigr)^{1/2}$, $\xi=k_1\mu_1\sec\theta_1 \tan\psi$. The matrix-elements
\begin{eqnarray}
 \alpha_l&=&  k_1\mu_2 \sec\theta_1 +\kappa \mu_1 \cos\psi+ \xi+i\, p\mu_1 \cr
 \beta_l &=& k_1\mu_2 \sec\theta_1-\kappa \mu_1 \cos\psi-\xi+i\, p\mu_1
\end{eqnarray}
and the transfer matrix $M_m$, that connects the electromagnetic fields at the far left ($z_l^+$) and far right ($z_r^-$) inside the metallic layer is written as mentioned before as
\begin{equation}
M_{m}=e^{-p d_c}\left( \begin{array}{cc} e^{i q d_c} & 0 \cr 0 & e^{-i q d_c} \end{array}\right).
\end{equation}
with an attenuation factor $e^{-p d_c}$, which depends on the attenuation constant $p$ and the conducting layer $d_c$, and a transfer matrix that accounts for the phases, $q d_c$ and $-q d_c$, gained by the electromagnetic fields that propagate along the  real angle $\psi$, with wave number $q$.

\begin{figure}
\includegraphics [width=15.6cm]{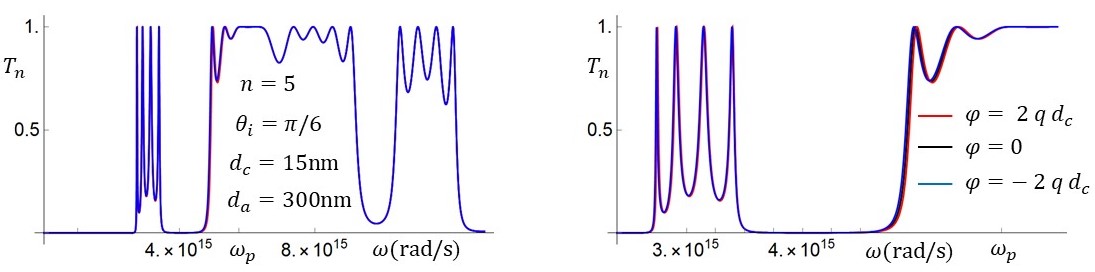}
\caption{The effect of  phase $\varphi$ in the transmission coefficients. The graph in (b) is a zoom of the graph in (a). The effect of the phase $\varphi$ is  negligible, and null for frequencies above $\omega_p$. }\label{f9}
\end{figure}

It is easy to verify that up to a phase $\varphi$, with negligible effects on the results as shown below, the requirement of flux conservation implies that
\begin{eqnarray}
s_R = \frac{\kappa \tanh{(p d_c)}}{\cos \psi} \hspace{0.3in}{\rm and }\hspace{0.3in}s_L= \frac{\kappa \tanh{(p d_c)}}{\cos \psi}e^{i(\pi+\varphi)}.
\end{eqnarray}
Since $s_R\propto\sigma$, we can define an absorption or flux-loosing factor $a$, such that $s_R=\sigma a$, thus the attenuation factor
\begin{eqnarray}
a = \frac{\kappa \tanh{(p d_c)}}{\sigma \cos \psi}\simeq \frac{\kappa \tanh{(d_c/\delta_c)}}{\sigma \cos \psi}
\end{eqnarray}
where $\delta_c$ is the skin depth, provides the magnitude of the induced currents and the amount of flux lost exciting the plasmonic resonances. Given the transfer matrix $M_{\jmath c}$ it is easy to obtain the unit-cell transfer matrix
\begin{eqnarray}
M_{\jmath}=\left(\begin{array}{cc}
          e^{i\,\theta_a/2}& 0 \\
          0 & e^{-i\,\theta_a/2}
        \end{array}\right) \left(\begin{array}{cc}  \alpha_{\jmath c} & \beta_{\jmath c} \\
                                        \gamma_{\jmath c} & \delta_{\jmath c}
                                      \end{array}  \right)\left(\begin{array}{cc}
          e^{i\,\theta_a/2}& 0 \\
          0 & e^{-i\,\theta_a/2}
        \end{array}\right)=\left(\begin{array}{cc} \alpha_{\jmath} & \beta_{\jmath}\\
                                        \beta_{\jmath}^* & \alpha_{\jmath}^*
                                      \end{array}  \right)
\end{eqnarray}
where $\theta_a=d_a k_1 \cos \theta_1$ and  $\alpha_{\jmath c}$, $\beta_{\jmath c}$,... are the matrix elements of $M_{\jmath c}$. Having the unit-cell transfer matrix, we can apply the theory of finite periodic systems.\cite{Pereyra1998,Pereyra2002,Pereyra2005,Pereyra2017} The transmission and reflection coefficients of a superlattice with $n$ unit cells, are obtained from
\begin{equation}\label{TransmTn}
  T_{\jmath n}=\frac{1}{|\alpha_{\jmath n}|^2}  \hspace{0.3in}{\rm and}\hspace{0.3in}   R_{\jmath n}=\frac{|\beta_{\jmath n}|^2}{|\alpha_{\jmath n}|^2},
\end{equation}
where
\begin{equation}\label{alfan}
  \alpha_{\jmath n}=U_n-\alpha_\jmath^* U_{n-1}  \hspace{0.3in}{\rm and}\hspace{0.3in} \beta_{\jmath n} = \beta_\jmath^{-1}U_{n-1},
\end{equation}
with  $U_n$ the Chebyshev polynomial of the second kind and order $n$, evaluated at the real part of $\alpha_{\jmath}$.
\begin{figure}
\includegraphics [width=15.6cm]{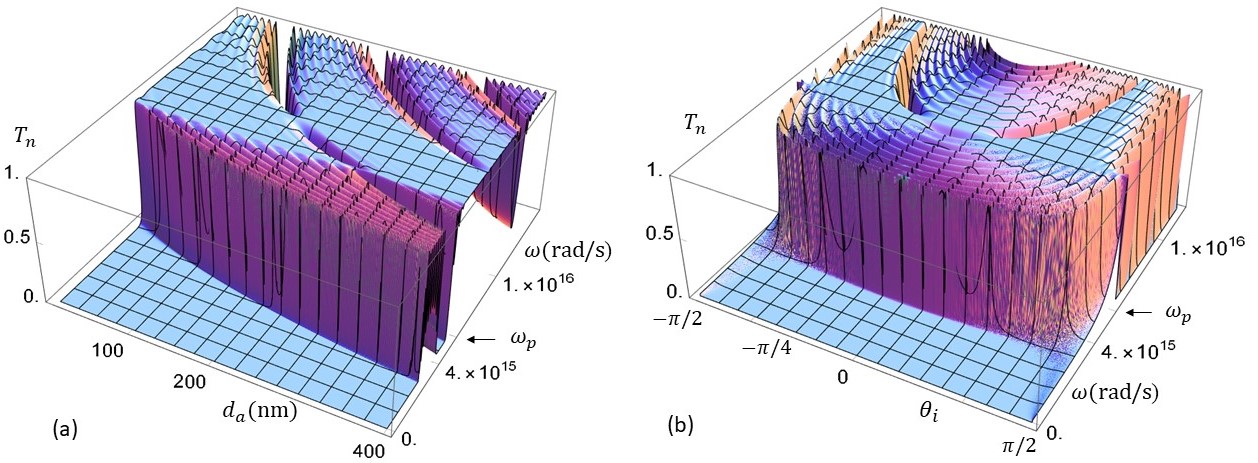}
\caption{Transmission coefficients as functions of  $d_a$ and $\theta_i$ plotted here to compare with those in figures \ref{f4} (a) and \ref{f6} (b). The agreement is perfect. }\label{f8}
\end{figure}
In figures \ref{f8} (a) and (b) we plot the transmission coefficients with the same parameters as in those of figures \ref{f4} (a) and \ref{f6} (b). The agreement below and above the plasma frequency is excellent. However, at $w_p$ and close to this frequency the behavior is different. For the graphs in Figure \ref{f8} we considered the phase $\varphi=0$. This phase accounts for the effective phase gain due to the multiple internal reflections. As shown in Figure \ref{f9}, the effect of $\varphi$ is practically negligible and null for frequencies above $\omega_p$. Figure \ref{f9} (b) is just a zoom of Figure \ref{f9} (a).
For the purpose of calculating only transmission and reflection coefficients there is no advantage using in this real-angle approach, compared with the more accurate complex-angle approach. However, the improved real-angle approach does inform on the role that the induced currents play in the energy-absorption phenomenon, for  frequencies below $\omega_p$, which implies the optical domain.

\begin{figure}
\includegraphics [width=15.6cm]{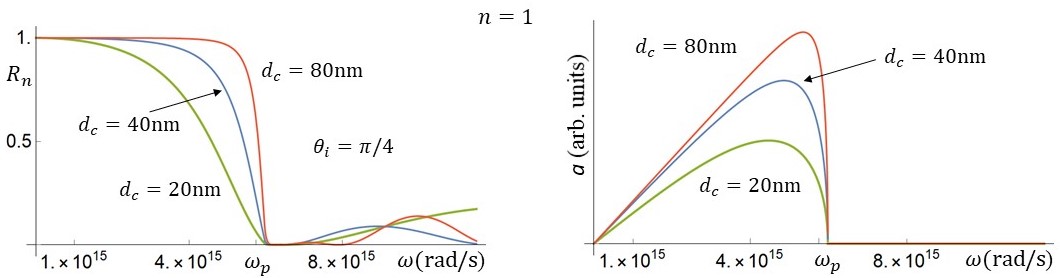}
\caption{Reflection and absorption factor for a single layer when the incidence angle is $\theta_i=\pi/4$. In (a), the reflection coefficient as function of the frequency, for different thicknesses of the silver layer. In (b), the absorption factor $a$ grows as the layers' thicknesses $d_c$ increase. The absorption factor vanishes at $\omega=0$ and $\omega=\omega_p$, and its maximum shifts to higher frequencies as $d_c$ grows. }\label{f10}
\end{figure}

\begin{figure}
\includegraphics [width=16.2cm]{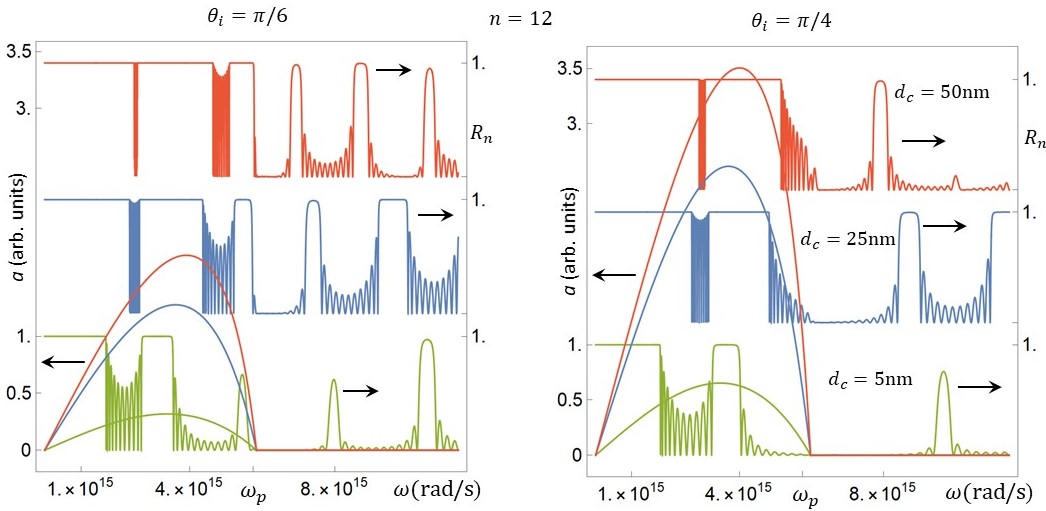}
\caption{Reflection and absorption factor for a superlattice with $n=$12, and two values of the incidence angle. The incidence angle at the left is $\pi/6$, and $\pi/4$ at the right. The bandwidths and the absorption factor grow with the incidence angle $\theta_i$ and the conducting layer thicknesses $d_c$.}\label{f11}
\end{figure}

In Figure \ref{f10} (a) we plot the reflection coefficient of a single layer of silver for different thicknesses, and in Figure \ref{f10} (b), the absorption factor $a$, which represents the strength of the absorbed flux exciting the plasmon oscillations. These currents attenuate the transmission and enhance the reflection coefficient. As shown in Figure \ref{f10} (b) the absorption factor $a$ grows when the conducting layer increases. In Figure \ref{f11} we show the effect of the incidence angle and conducting layer width on the absorption factor $a$ and on the reflection coefficient $R_n$ for a superlattice with $n=12$ and dielectric width $d_a=400$nm. As shown by these results the absorption factor is very sensitive to the incidence angle and, of course, to the conducting layer width. We see that increasing the incidence angle from $\theta_i=\pi/6$ to $\theta_i=\pi/4$ the factor $a$, hence the induced currents, grows by a factor of 2.
Before we conclude this section, let us see how is the resonant dispersion relation, and particularly how are the dispersion relation  predictions below the plasma frequency $w_p$. As mentioned before the resonant dispersion relation derived in the theory of finite periodic systems is\cite{Pereyra2005,Pereyra2002,Pereyra2017,Pereyra2020}
\begin{equation}
\cos \frac{\nu+(\mu-1)n}{n}\pi=(\alpha_{\jmath R})_{\mu,\nu}\hspace{0.3in} {\rm with} \hspace{0.3in}\mu=1,2,3,...  \;\; \nu=1,2,...,n-1.
\end{equation}
In terms of the physical quantities defined in this approach, this relation (for $\varphi =0$) becomes
\begin{eqnarray}
\cos \frac{\nu \!+\!(\mu\!-\!1)n}{n}\pi\!=\left(\cosh{p{d_c}} \cos qd_c \cos \theta_a -f_1\cosh{p{d_c}} \sin qd_c \sin \theta_a+f_2\sinh{p{d_c}} \cos qd_c \sin \theta_a \right)_{\mu,\nu}
\end{eqnarray}
with
\begin{eqnarray}
  f_1 &=&\frac{\mu_2 \sin 2\psi}{2 \mu_1 \sin 2\theta_1} \left(1 +\frac{p^2 \mu_1^2}{k_1^2\mu_2^2}\cos ^2\theta_1 +\frac{\mu_1^2 \sin^2 2 \theta_1}{\mu_2^2 \sin^2 2\psi} \right)\cr \cr
  f_2 &=&\frac{\mu_1 \sin 2\theta_1}{2\mu_2 \sin 2\psi} \left(\tan qd_c\tanh pd_c+\frac{p}{ k_1}\frac{\sin 2 \psi}{\sin \theta_1}  \right)
\end{eqnarray}
A quantity that is also useful is the imaginary part of $\alpha_{\jmath}$, that can be written as
\begin{eqnarray}
\alpha_{\jmath I}=\cosh{p{d_c}} \cos qd_c \sin \theta_a + f_1\cosh{p{d_c}} \sin qd_c \cos \theta_a-f_2\sinh{p{d_c}} \cos qd_c \cos \theta_a.
\end{eqnarray}
\begin{figure}
\includegraphics [width=15.6cm]{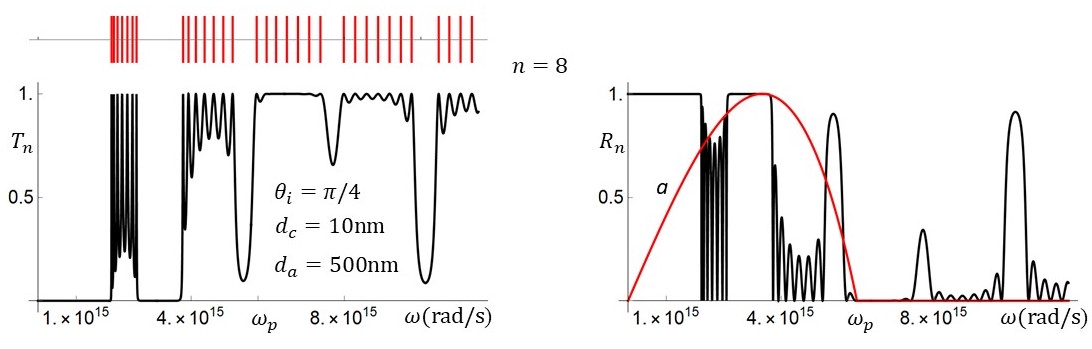}
\caption{Transmission and reflection coefficients for a superlattice with $n=$8, $d_a=50$nm, $d_c=10$nm, and incidence angle $\theta_i=\pi/4$. In the left, upper part, we have also the predicted resonant frequencies from the dispersion relation. In the right the absorption factor. }\label{f12}
\end{figure}
In Figure \ref{f12} we plot the transmission and reflection coefficients, together with the absorption factor and the resonances and bands predicted by the dispersion relation.
These graphs show not only the ability of this approach to calculate the essential quantities like the transmission and reflection coefficients, and the accurate prediction of the resonance frequencies spectrum, it shows also the appropriateness of this approach to determine, through the absorption factor $a$, the strength of the absorbed energy consumed to excite the plasmonic polaritons as response to the incident electromagnetic field. This information provides additional insight in the physics of the metallic superlattices.

In the applications based on the response of metallic-structures to electromagnetic fields, the response to electromagnetic pulses might be of interest. Since the details of the electromagnetic fields inside the metallic structure is a bit complex, we will just present here a couple of results.

\section{Reflection and transmission of Gaussian pulses by metallic superlattices}

As illustrative examples of use of the above mentioned results, we will present the transmission and reflection of an electromagnetic pulse by a metallic superlattice.  If the electromagnetic pulses are Gaussian wave packets defined by
\begin{eqnarray}
\Psi_E(x,z,t,\theta_i)=\int_{-\infty}^{\infty}dk e^{-\gamma(k-k_o)^2}e^{i(\bf{k\cdot r}_o -\omega t)}{\bf {\cal E}}(k,z,\theta_i)=\int_{-\infty}^{\infty}dk e^{-\gamma(k-k_o)^2}{\bf  E}(k,z,\theta_i,t)
\end{eqnarray}
\begin{figure}
\includegraphics [width=15.6cm]{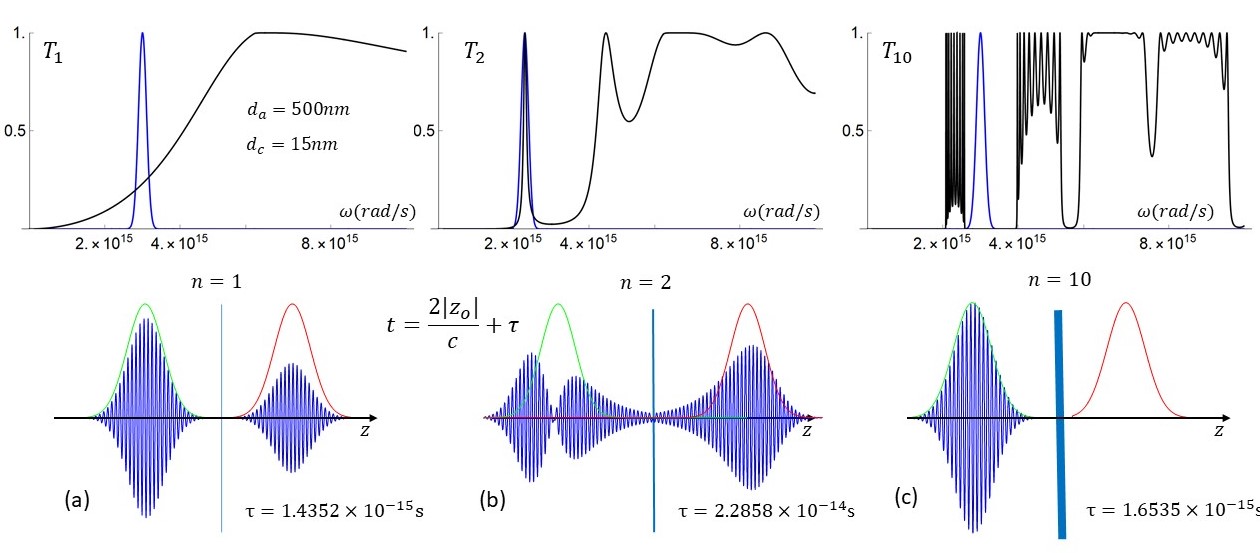}
\caption{Space time evolution of a Gaussian wave packet through metallic superlattices with $n=$ 1, 2, and 10, in the left, middle and right, respectively. In the upper panels we show both the transmission coefficients and the Gaussian wave packets at the frequencies at which they were defined at $z_o$ and $t=$0. In the lower panel, we have the reflected and transmitted wave packets at $t=2 |z_o|/c+\tau$, where $\tau$ the tunneling time of the wave packet components of frequency $\omega_o$. The green and red Gaussian curves, define the positions at which the reflected and transmitted packets should be if $\tau$ is truly the tunneling and reflecting time. Notice that the reshaping of the transmitted and reflected packets are strongly determined by the transmission coefficients in the frequencies domain where the packets are defined, at a resonance in the middle and at a gap in the right.}\label{f13}
\end{figure}where $k_o$ defines the peak of the Gaussian pulse and $\bf{r}_o$ its position at $t=0$.  For $z<0$, the $z$ component of the electromagnetic field is written as
\begin{eqnarray}
 E_z(k,z,t,\theta_i)={\cal E}_o\left( e^{i k \cos \theta_i(z+z_o-v_gt)}-\frac{\beta_n^*}{\alpha_n^*}e^{-ik\cos \theta_i(z-z_o+v_gt)} \right) \hspace{0.3in} z<0
\end{eqnarray}
where $\alpha_n$ and $\beta_n$ are the matrix elements of the $n$-unit cells transfer matrix $M_n$.  For $z>L=nl_c$, with $l_c$ the SL unit cell length, the electromagnetic field is written as
\begin{eqnarray}
 E_z(k,z,t,\theta_i)={\cal E}_o\left(\alpha_n-\beta_n\frac{\beta_n^*}{\alpha_n^*}\right)e^{i k\cos \theta_i(z-L+z_o-v_gt)}\hspace{0.3in} z>L
\end{eqnarray}

The possibilities of total, partial or zero reflection of an electromagnetic pulse by metallic superlattices are important for applications and strongly determined by the transmission or reflection coefficients of each component of the Gaussian packet, in particular by the domain of frequencies where the wave packet is defined. Hence, the transmission coefficients for the Gaussian pulse components are different. This difference manifests also in the shape of the reflected and transmitted pulses, as shown in the lower panels. In the figure we plot the $z$-components as functions of $z$ only. We  assume also that the fields are in parallel polarization and the incidence angle is $\theta_i=\pi/4$.

To visualize the effect of the metallic SL on a Gaussian pulse we consider three cases where the Gaussian packets are similar, but the characteristics of the domain of frequencies where they are defined are different, and the number of unit cells in the metallic SLs is different. The number of unit cells  in (a), (b) and (c), is 1, 2 and 10, respectively.

In figure \ref{f13} the transmitted and reflected wave packets are shown at $t=2|z_o|/c+\tau$, assuming that the fields outside the SL move with velocity $c$, and the centroid with wavenumber $k_o$ spends a time equal to phase time $\tau$ inside the SL.\cite{Spielmann1994,Pereyra2000} In the lower panels we have also the envelopes of the wave packets at $t=0$, green, (in position $-z_o$) and at $t=2|z_o|/c+\tau$, red.  The green and red Gaussian curves define the positions at which the reflected and transmitted packets should be found if $\tau$ is truly the tunneling or reflecting time. In each case, the tunneling time $\tau$ of the packet peaks is indicated. The prediction is correct, as was shown in other papers.\cite{Simanjuntak2007,Pereyra2011} In case b), and slightly in case a), the asymmetry of the tunneling times implies that the components of one tail (those with smaller frequency) move faster than those in the other tail. For this reason, in order to plot the whole transmitted and reflected packets in case b), we increase the distance $|z_o|$.

Since, the scattering process is different for each wave packet component $E(k,z,\theta_i)$, the wave packets are, generally, distorted, unless the whole wave packet is defined in a frequency domain such that almost all components have the same transmission coefficients. This is the case  in figure \ref{f13} (c), where the wave packet is defined in a gap, with $T_n\simeq$0, being thus almost completely reflected.   In figures \ref{f13} (a) and (b), because of the frequency domains, the wave packet in a)  is partially transmitted and partially reflected. The wave packet in b), defined at a resonance, is strongly distorted. The components close to the Gaussian peak are transmitted while those in the tails are reflected.

\section{Conclusions}
We have shown that the anomalous results and apparent parity effects reported in Ref. [\onlinecite{Pereyra2020}], are consequence of the common assumption that electromagnetic fields move along the direction of propagation of the constant-phase planes, the finiteness requirement, and the neglection of the induced currents. We have shown that this assumptions imply a lack of unitarity related to the underlying phenomena of absorption and loss of energy. To cure this problem, we introduced two approaches. On the one hand, we have shown that by keeping the complex angles, the principle of flux conservation is fully satisfied, above and below $\omega_p$. The anomalous results for frequencies below $\omega_p$ disappear, and the predictions above $\omega_p$ obtained in Ref. [\onlinecite{Pereyra2020}], remain. The complex-angle approach presented here, preserves all the information of the scattering process in the metallic superlattice, and gave us light to improve the formalism when the real angle assumption is made. In fact, we have shown here that taking into account the induced currents and the requirement of flux conservation, we end up with an improved approach, with new Fresnel and transmission coefficients, fully compatible with those of the complex-angle approach. The improved approach allows also to evaluate the magnitude of the induced currents and the absorbed energy, as functions of frequency and the superlattice parameters. We determine the plasmonic resonant frequencies, and present preliminary results of the response of metallic superlattices to electromagnetic pulses and wave packets, particularly, in the optical domain. We calculate the reflection and transmission coefficients as well as the phase time $\tau(\omega)$. We show that the predicted space-time positions agree extremely well with the actual positions of the wave-packet centroids.

\section{Acknowledgements}
The author acknowledges comments of H. P. Simanjuntak.


\begin{thebibliography}{99}
\bibitem{PlasmaFreq} Since the plasma frequency is the frequency obove which the real part of the dielectric function becomes positive, the plasma frequency of silver is $E_p/\hbar\sqrt{\epsilon_{\infty}}=9\times10^{16}eV/(6.5821\sqrt{5.7}=5.727\times10^{15}s^{-1}$.
\end{thebibliography}
\end{document}